%% file: draft3.tex
\documentclass{sigchi}
\pdfoutput=1

\usepackage{balance}  
\usepackage{graphicx} 
\usepackage{times}    
\usepackage{url}      

\usepackage{amsmath}
\usepackage{amssymb}
\usepackage{amsfonts}
\usepackage{wrapfig}
\usepackage{graphicx}
\usepackage{color}
\newcommand{\bi}[1]{\mathbf{#1}}

\makeatletter
\def\url@leostyle{%
  \@ifundefined{selectfont}{\def\UrlFont{\sf}}{\def\UrlFont{\small\bf\ttfamily}}}
\makeatother
\def\pprw{8.5in}
\def\pprh{11in}

\usepackage[pdftex]{hyperref}
\hypersetup{
pdftitle={SIGCHI Conference Proceedings Format},
pdfauthor={LaTeX},
pdfkeywords={SIGCHI, proceedings, archival format},
bookmarksnumbered,
pdfstartview={FitH},
colorlinks,
citecolor=black,
filecolor=black,
linkcolor=black,
urlcolor=black,
breaklinks=true,
}

\toappear{}

\begin{document}

\title{SurfCuit: Surface Mounted Circuits on 3D Prints}

\author{Nobuyuki Umetani\;\;\;\;\;\;\;\;\;\;\;Ryan Schmidt\vspace{9pt}\\
    \affaddr{Autodesk Research}\\
    \email{\{nobuyuki.umetani,ryan.schmidt\}@autodesk.com}\\
}

\maketitle

\begin{abstract}
We present, SurfCuit, a novel approach to design and construction of electric circuits on the surface of 3D prints.
Our surface mounting technique allows durable construction of circuits on the surface of 3D prints.
SurfCuit does not require tedious circuit casing design or expensive set-ups, thus we can expedite the process of circuit construction for 3D models.
Our technique allows the user to construct complex circuits for consumer-level desktop fused decomposition modeling (FDM) 3D printers.
The key idea behind our technique is that FDM plastic forms a strong bond with metal when it is melted.
This observation enables construction of a robust circuit traces using copper tape and soldering.
We also present an interactive tool to design such circuits on arbitrary 3D geometry. 
We demonstrate the effectiveness of our approach through various actual construction examples.
\end{abstract}


\category{H.5.m.}{Information Interfaces and Presentation (e.g. HCI)}{Miscellaneous}

\input{intro_related}

\input{method_construction}

\input{method_software}

\input{result}

\section{Future Work}

%
%
%

We are interested in making the circuit design system more intelligent by incorporating a circuit simulator (e.g., SPICE), a physics engine (e.g., Open Dynamics Engine) to simulate printed characters' dynamics, a schematic image recognition system~\cite{Arvo:2000:FSC:354401.354413}, or an interactive sketch beautification system~\cite{Igarashi:1997:IBT:263407.263525} to facilitate the user's creative circuit integrated 3D object design.

\section{Conclusion}
We presented SurfCuit: a system that integrates circuits into 3D prints by mounting them on the printed surface.
Our construction method enables building rather complex, highly-conductive circuit patterns robustly on FDM-based 3D prints.
Our interactive design system enables intuitive input and 3D layout of electric circuits on 3D geometry.

%
%
%
%
%
\balance

\bibliographystyle{acm-sigchi}
\bibliography{sample}
\end{document}

%% file: intro_related.tex
\section{Introduction}
\label{sec:introduction}

Recent advances in consumer 3D printing technology have made it possible for end users to casually fabricate 3D plastic objects.
Many 'makers' would like to add interactivity to their printed objects using sensors, lights, motors, and so on.
However, incorporating the necessary electric circuits into these objects has not become inherently easier with 3D printing.
Electric circuits are typically designed in 2D, and mounted on planar geometries, such as printed circuit boards (PCBs)
Inserting a flat circuit inside a 3D object requires extensive geometry editing to create cavities, wire routing paths, and fixtures.
This is generally beyond the reach of the novice or casual maker.

\begin{figure}[t!]
\includegraphics[width=86mm]{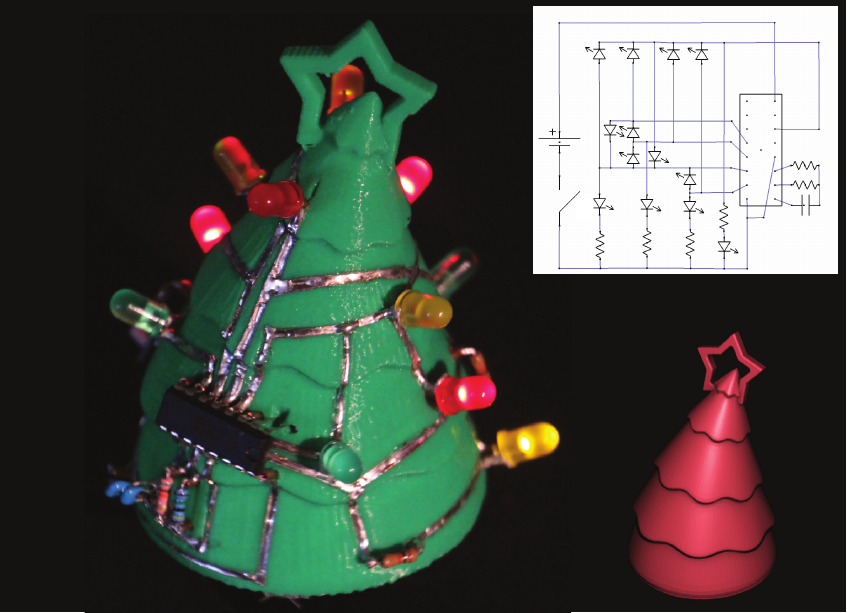}
\caption{SurfCuit allows the user to design and fabricate surface mounted circuits on 3D prints (left). An illumination circuit (top right) is mounted on the surface of Christmas tree shape (bottom right). }
\label{fig:teaser}
\end{figure}

An alternative to is to use 3D circuitry, where 3D traces are embedded into the object volume or surface.
However existing CAD interfaces and fabrication techniques have not been designed with 3D circuits in mind.
In this paper, we demonstrate both a design tool and fabrication technique to integrate the mechanical and electrical functions of simple objects.
Our approach, which we call SurfCuit, allows the user to design and construct functional and durable electric circuits on the surface of 3D prints (see Fig.~\ref{fig:teaser}).

As a prototyping method, surface mounting has various advantages over embedded circuitry.
First, construction is much easier since the parts are accessible from outside -- the user does not need to insert the electric parts during printing, or try to fit components into tiny cavities.
Second, it is easy to debug and repair surface-mounted circuits, while in many embedded-circuit applications this is difficult or impossible.
Finally, the circuit design task is greatly simplified -- compared to three-dimensional arrangements of cavities, fixtures, and wire channels, surface layouts are intuitive and efficient to create.

Our key challenges are (i) how to fabricate complex circuits on the surface of 3D prints and (ii) how to help the user perform the \emph{circuit layout} task directly on the 3D surface in a computational design tool. 
Conductive inks do not readily adhere to 3D print plastics and their resistance is too high for many types of components.
Instead, SurfCuit uses copper tape and tubes that are soldered together to achieve mechanically durable and highly conductive circuits. 
We leverage the fact that near soldering temperature, 3D-printed PLA plastic melts to a sticky viscous fluid that bonds well to copper material.
To further enhance the fabrication process, our design tool adds shallow channels and holes to the 3D model before printing.
These channels both help the user to re-create their virtual circuit traces in the physical world, and also help to firmly affix the copper tape and through-hole parts onto the 3D print.

The traces which connect components in an eletrical circuit must be physically isolated, i.e. they cannot intersect with eachother.
Because they exist in a (possibly curved) 2D space, with even a moderate number of traces it becomes very difficult 
to create the circuit without carefully planning out the traces ahead of time.
Board layout and planning software like EAGLE is an essential tool for planar circuit desgin.
However no existing circuit planning tool is applicable to arbitrary 3D surfaces.
In addition, strips of copper tape cannot follow arbitrary paths on 3D surfaces, as many paths 
introduce too much torsion into the tape, which will result in kinks or tears.
The strips should be laid out along \emph{geodesics}, and without computational 
guidance it is very difficult to ensure that the 3D traces have this property.
SurfCuit offers an interactive design tool that allows the user to easily adapt an existing planar circuit schematic to a 3D surface.

To demonstrate the capabilities of SurfCuit, we have designed and fabricated a variety of 3D objects with 3D circuitry.
Some of these examples involve circuit complexity and levels of voltage/current that are well beyond what has been demonstrated in the literatuer.
In addition, we also perform some destructive testing to demonstrate the robustness of our fabrication process.
We show various examples of how our system facilitates the user's creation of functional 3D objects with electric circuits. 

\section{Background and Related Work}
\label{sec:related work}

\subsection{Molded Intergrated Devices}
\begin{figure}[htbp!]
\centering
\includegraphics[width=86mm]{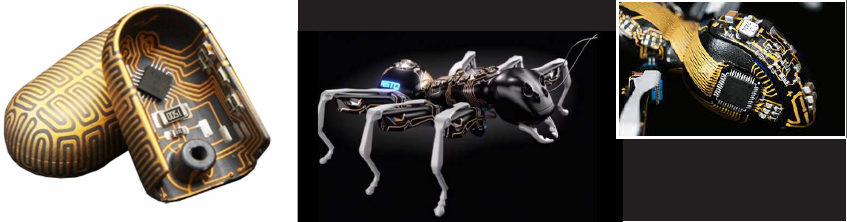}
\caption{Example of molded interconnected devices. (Left) A robotic finger tip sensor by CITEC, Bielefeld University. (Right) FEST Bionic Ant Robot}
\label{fig:MID}
\end{figure}

Various advanced manufacturing technologies support the fabrication of circuitry that conforms to curved surfaces.
For example, the Optomec Aerosol Jet system can create metal traces on simple 3D forms using laser metal deposition techniques.
Another technology, Molded Interconnect Devices~(MID), makes it possible to install circuitry on plastic surfaces.
MID enables functional integration of electric circuit into small spaces, thus is often used for compact implementation of electronic products such as cell phones, cars, and advanced micro robots (see Figure~\ref{fig:MID}).
%
%
However, manufacturing such objects involves multi-axis laser engraving machines and etching/plating baths only found in advanced industrial facilities.
Our SurfCuit system is inspired by these techniques, but our goal is to introduce MID-style fabrication in a more accessible context.
In addition, currently there are no CAD tools for MID design.
Our SurfCuit design tool is directly applicable to these advanced manufacturing methods.

\begin{figure*}[t!]
\includegraphics[width=177.8mm]{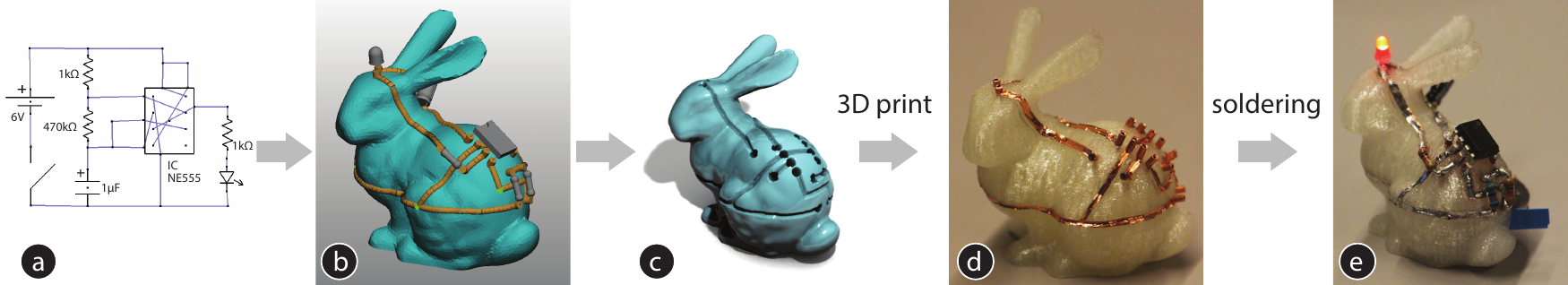}
\caption{Workflow of SurfCuit. The user first draws a 2D schematic diagram of a circuit (a), then positions the electrical components on a 3D shape and connects them with curved traces (b). 
SurfCuit automatically generates channels and holes on the surface (c) to guide the user in placement of copper tapes and tubes. 
Finally, the user solders the copper pieces together to achieve a robust circuit on the 3D print.}
\label{fig:workflow}
\end{figure*}

\subsection{Interactive 3D Prints}
Various works in the human-computer interaction and fabrication literature have addressed the topic of adding interactivity to 3D prints.
Techniques have been presented to convert 3D prints into sensors that include cameras~\cite{Savage:2013:SES:2501988.2501992}, acoustics~\cite{Ono:2013:TAA:2501988.2501989} and light-guides~\cite{Willis:2012:POP:2380116.2380190}. 
Sato et al.~\cite{Sato:2012:TET:2207676.2207743} studied frequency-dependent impedance properties to detect configurations of conductive 3D objects.
Printput~\cite{raey} and Capricate~\cite{Schmitz:2015:CFP:2807442.2807503} use conductive filaments to convert the surface of 3D prints into capacitive touch sensors. 
However, such conductive filament traces have very high resistance, making it difficult to supply enough current to drive larger components (See Section~Resistance Comparison).

Fabricating 3D circuit traces inside "tubes" passing through the interior of 3D prints was explored by Savage et. al.~\cite{Savage:2014:STA}.
This method is effective at hiding the circuitry, but also illustrates the challenge of repairing such devices.
The capabilities of the automatic routing algorithms also limit the complexity of the circuits that can be designed with this approach.
Hudson~\cite{Hudson:2014:PTB:2556288.2557338} studied the use of conductive threads for the yarn-based soft 3D printing.
The commercially-available Voxel8 3D Printer~\cite{voxel8} can embed circuits in a plastic print using conductive inks.
The liquid ink can be deposited on the outer surface of prints, but only in upward-facing regions.

\subsection{Prototyping Flat Circuits}
Solderless prototyping methods such as traditional breadboards and LittleBits~\cite{Bdeir:2009:EML} are the fastest way to 
assemble a circuit, but these methods not suitable for permanent use. 
The resulting circuits are fragile, and also space consuming.
Recently, advances in conductive ink have made it possible to directly deposit circuit traces on flat sheets
using consumer ink-jet printers~\cite{Kawahara:2013:IIC}.
ShrinkCuit~\cite{Lo:2014:SSS:2642918.2647421} uses Shrinky Dinks\texttrademark~ as the substrate to enhance 
complexity and conductivity of conductive-ink-based circuits.
Sketch In Circuit \cite{Qi:2014:SCD:2556288.2557391} uses copper tape for prototyping traces on paper.
Circuit Sticker~\cite{Hodges:2014:CSP} enables rapid prototyping of more complex circuits by pasting ready-made circuit boards on the top of traces.
Ramakers et al.\ presented an interactive design system for circuits printed on paper~\cite{Ramakers:2015:PIA:2702123.2702487}.
The motivation behind these works is to enable creative circuit prototyping for ``makers" and non-experts.
We share this motivation, and our goal is to extend these ideas to arbitrary free-form 3D surfaces with a robust construction technique and interactive trace layout interface.

%

\subsection{Circuit Design Tools}
There are many circuit layout design tools for planar circuit boards.
For example, commercial packages such as Eagle\texttrademark, AutoTRAX\texttrademark and DipTrace\texttrademark provide comprehensive environments for design and simulation for PCB. 
Autodesk 123D Circuits~\cite{123Dcircuits} provides a layout design system for breadboards.
Autodesk Project Wire provides 3D circuit layout for the Voxel8 printer~\cite{voxel8}.
Our SurfCuit 3D circuit design tool enables novices to quickly design 3D traces constrained to the surface of an arbitrary input mesh.


%% file: method_construction.tex

\section{Surfcuit Circuit Fabrication}
\label{sec:implementation}

\subsection{Construction Procedure}
The robustness of electrical connections is very important for permanent circuit construction --- disconnection of a single trace can disable an entire circuit.
However, constructing robust, highly conductive traces on a curved 3D-printed surfaces has been difficult.
In this paper, we take advantage of the fact that PLA and ABS plastics that they melt into sticky, 
glue-like viscous fluids at around 200$^\circ$C.
Since the melting point of solder is also around that temperature, soldering on traces 
and pins melts the surrounding 3D plastic and strengthens the mechanical bonds between them.
The result is that after cooling, the highly conductive copper traces are firmly affixed to the surface of the 3D print.
Note that our approach is inspired by recent works that exploit melting 
behaviors for fabrication~\cite{Mueller:2014:LOL:2590181.2567782,Sageman-Furnas:2015:MFC:2820903.2820915}. 
However here we use melting for bonding, not for forming.

This melting technique creates strong connections, but actually using it on 3D surfaces requires some pre-planning.
Computational design tools are necessary to plan the spatial layout for complex circuits.
To guide the user in fabricating their design, we computationally generate channels and 
holes in the 3D surface before printing, which also helps to increase circuit robustness.
Hence, the workflow of SurfCuit fabrication is as follows (see Fig.~\ref{fig:workflow}): 
\begin{enumerate}
\item In the SurfCuit design tool, the user first draws a 2D circuit schematic diagram, then lays out the electric parts and traces on a 3D surface. 
\item SurfCuit generates a 3D mesh with channels and holes that guide the user to install copper tape and tubes on a 3D print. 
\item Then, the user covers these copper traces with solder, and solders the traces, pins, and electric parts together.
\item Finally, a thin layer of clear lacquer spray electrically protects the traces.
\end{enumerate}

With SurfCuit, a novice maker can easily create complex functional 3D objects using widely-available single-material FDM printers.
The soldering process requires exactly the same skills as fabricating a 2D circuit, which we already know that nearly anyone can learn to do. 
Applying solder to the copper tape is not difficult, as the solder naturally flows into the trace channels due to surface tension (see accompanying video).  
This soldering step also thickens the traces, further increasing mechanical robustness and electrical conductivity.

\subsection{Construction Detail}

\begin{figure}[htbp!]
\includegraphics[width=86mm]{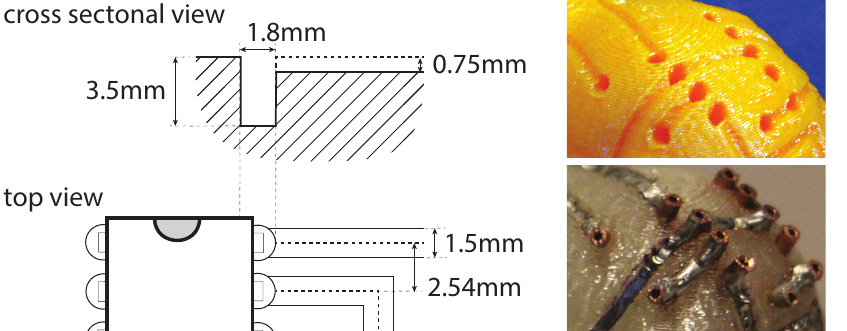}
\caption{(left)~Dimensions of holes and channels generated by our tool to guide copper tape and tube placement. 
We used 1.5~mm-wide copper tape and copper tubes with 1/16~inch (1.6~mm) diameter. 
The images on the right show the channels and holes in a 3D print (top), and the post-soldering result (bottom). }
\label{fig:construction}
\end{figure}

Our circuit fabrication process is intended to be used with though-hole parts.
Though-hole parts are desirable because they are widely available and are easy to manually solder. 
More importantly, through-hole parts achieve mechanically stronger bonds compared to surface mounted parts. 
Thus, they can be left exposed on the surface.
The interval between pins is typically 1/10~inch  = 2.54~mm for though-hole parts. 
Our technique maintains sufficient isolation between neighboring traces by keeping the trace width 
and pin diameter smaller than this interval (see Fig.~\ref{fig:construction}).

The channels and holes generated by our design tool are essential for the fabrication process.
We arrived at this process after multiple iterations with simpler techniques.
The benefits of this method include:
\begin{itemize}
\item The channels and holes help the user to accurately re-produce their complex virtual designs
\item The channels and holes provide a large contact area between the copper traces and the 3D print.
\item The holes help to temporarily hold components in place while they are soldered
\item The channels partially enclose the traces and prevent them from peeling off.
\item Since copper tape is placed in a V-shaped channel, it is easy to cover the channel with solder -- surface tension makes the solder naturally flow into the channel. 
\end{itemize}

The use of solid copper, rather than conductive inks or similar alternatives, is desirable for three reasons.
First, copper has comparatively high conductivity / low resistance (less than $0.5~\Omega/\mathrm{m}$ in our traces). 
Thus, we are not limited by parasitic resistance when using the large electric currents typically necessary for driving small motors or incandescent lights.
Secondly, its high heat conductivity makes it possible to bond copper to 3D-printed plastics via the application of heat.
Finally, the copper is solderable. 
Solder quickly spreads on the surface of copper and creates mechanically and electrically robust bonding.

Compared to silver-based conductive inks (the most conductive), copper is also inexpensive and widely available.
To make 1/16 inch traces, we split widely available 1/8 inch copper tape (about 10\$ for 50~m) in half using a rotary cutter. 
For copper pins, we used 1/16 inch copper tubes manufactured via K\&S Engineering .Inc (three dollars for 1~m).

\subsection{Resistance Comparison}
During the development of our fabrication process, we experimented with various conductive inks and copper paints.
However we found that most are difficult to apply to plastic surfaces.
Liquid-based materials were also unreliable as a small crack results in electrical disconnection.
This is especially problematic on the rough surfaces of 3D FDM prints, where the ink will tend to pool into the small cavities and channels produced by the printing process, creating highly variable thickness in the conductive layer.
Furthermore, as reported in the previous studies, the electric resistance in the conductive liquids are very high (e.g,  11.48~$\Omega$ for a 28cm~x~0.5cm trace using silver conductive ink~\cite{Lo:2014:SSS:2642918.2647421} and 2~$\Omega$/inch for a~ 3mm diameter tunnel filled with copper paint~\cite{Savage:2014:STA}).
%

We also tried drawing traces with conductive PLA filament using a hand-held plastic extruder, 
however again the resistance is significantly higher than typical PCB traces, and thus cannot support many common circuits. 
For example, Black Magic 3D's Conductive Graphene Filament, which has one of the highest conductivity among the 
filaments on the market, still has 0.6~$\Omega/\mathrm{cm}^3$ volume resistivity.
To achieve similar resistance as our traces (0.5~$\Omega/\mathrm{m}$), 
the cross-sectional area should be at least 120~$\mathrm{cm}^2$(=11cm~x~11cm), vastly larger than the through-hole component pitch (2.54~mm).
In other words, if we use conductive filament for traces, and the traces have a 1.5mm~x~1.5mm cross-section (small enough to connect to through-hole components), then a 20~cm long trace has more than 0.5~k$\Omega$ resistance.
Such resistance causes over a 1V voltage drop with only 2mA current, which is barely enough to light an LED.
Generally, conductive filaments are sufficient for capacitive touch sensors or blinking LEDs, 
however they fall short when attempting to drive common micro-controllers, actuators, and transducers.

\subsection{Robustness Comparison}
\begin{figure}[htbp!]
\includegraphics[width=86mm]{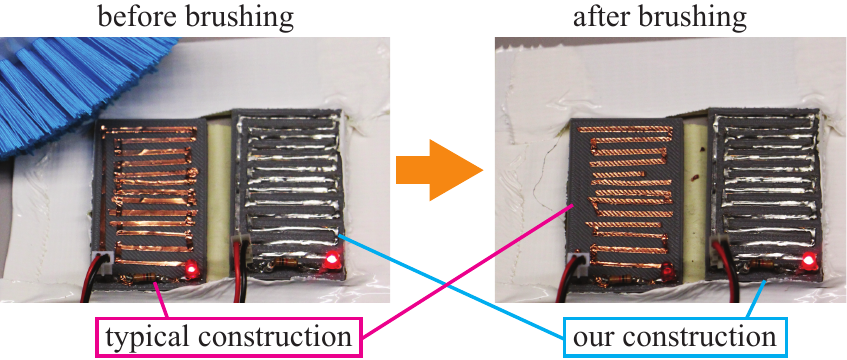}
\caption{(left)~LED lighting circuits with na\"ive construction and SurfCuit construction. 
(right) After brushing for one minute, the traces in the na\"ive construction were severely damaged while the SurfCuit traces were intact.}
\label{fig:robustness}
\end{figure}

The user of copper tape in circuits-on-surfaces is not entirely novel, in particular copper tape is widely used in wearables and fashion tech.
However these circuits are generally very fragile. 
To demonstrate the mechanical robustness of our approach, we made a qualitative comparison 
with a na\"ive method using destructive testing (see Figure~\ref{fig:robustness}).

In the na\"ive construction method, traces are just copper tapes placed on the 3D prints without the channels and solder.
The tape generally has a sticky backing which is sufficient to hold it in place.
The test circuits light a LED using a trace pattern that consists of more than twenty copper tapes (see Figure). 
We then brushed the circuits for one minute with nylon brushes to observe the robustness of the circuit.
We first used a relatively soft nylon kitchen brush designed for washing dishes, and then switched to very hard nylon brush meant for scraping off rust.

The circuit with na\"ive construction immediately stopped functioning when the soft nylon brush was applied.
This is because the connection between overlapping copper tape segments is particularly weak and breaks under small mechanical forces.
After few additional seconds of brushing, the copper tape segments in the na\"ive construction start to peel off the plastic.
After one minute, the copper traces in the na\"ive construction were severely damaged, to the point where the circuit would need to be entirely rebuilt.
On the other hand, there was no visual damage to the SurfCuit traces even after the brushing with the hard brush.
The circuit with SurfCuit construction stayed functional during the entire experiment.
Please refer to the accompanying video for the detail of the comparison.

%% file: method_software.tex
\section{SurfCuit Design Tool}
Designing the trace layout for a complex circuit (e.g., 10+ connection points) typically requires planning before starting construction.
Each trace that is added constrains the design space of future traces, because no traces can intersect.
For the 2D flat circuits, we can plan on a paper or in vector-graphics CAD tools.
However, since our circuits are on 3D free-form surfaces, we cannot lay out traces on a 2D flat geometry.
It is also not practical to lay out traces as 3D space curves, as keeping them "on the surface" is very cumbersome.
Hence, we developed a domain-specific interactive CAD tool which allows users to arrange parts and traces on arbitrary 3D surfaces.

\subsection{User interface of Surfcuit design tool}
\begin{figure}[htbp!]
\includegraphics[width=86mm]{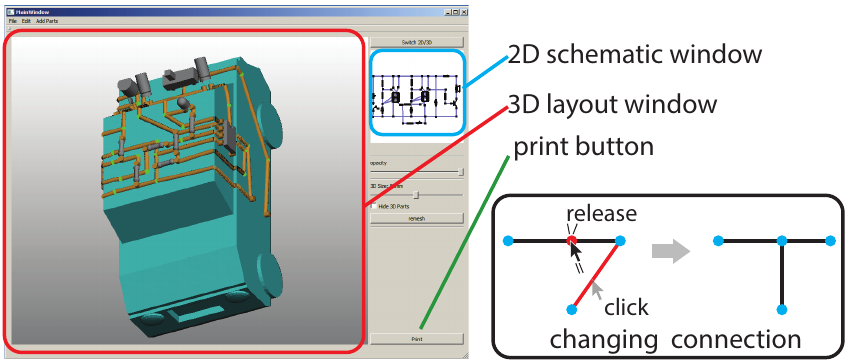}
\caption{(left)~SurfCuit 3D trace design interface. The user designs traces on a 3D object by dragging and rotating parts. 
The 2D schematic window is shown at the same time to facilitate understanding of the circuit 
structure. (right bottom)~The user can change connectivity of traces by a simple gesture.}
\label{fig:interface}
\end{figure}

Our SurfCuit design tool has two modes: 2D schematic design mode and 3D part and trace layout mode (see Fig.~\ref{fig:interface}). 
The schematic mode lets the user specify electronic parts and their connections in the form of a 2D diagram, while the 3D layout mode lets the user arrange the parts and traces on the 3D model's surface.
The user can switch back and forth between these modes during circuit editing. 
While the user edits the circuit in one mode, a small window highlights the electric parts or traces currently being edited in the other mode, making 2D/3D correspondence easy to understand.
Note that we are inspired by existing works showing highly abstracted schematic diagram while editing complex models \cite{Zhu:2011:SDI:2070781.2024168,Kazi:2014:KSD:2642918.2647375}.

The schematic diagram is desirable as a circuit input rather than drawing traces directly on a 3D surface because schematics are symbolized and thus easily comprehensible.
In our schematic editor, the user places symbols of electric parts and specifies these connections.
To make the schematic tidy, the user also can add or delete a point on a trace and switch connectivity of points inside connected traces (see Fig.~\ref{fig:interface}-right bottom).
%
%
Note that thousands of schematic diagrams are widely available on the internet, for virtually any kind of circuit. 
Thus, inputting such diagrams does not require sophisticated electronics knowledge, a novice designer can simply copy an existing schematic.

The electric parts and their connections, specified in the 2D schematic window, are imported to the 3D layout window.
The user lays out the parts by dragging and rotating them on the 3D surface.
The traces are automatically generated on the surface in real-time during editing, making it easy for the user to lay out parts while avoiding intersecting traces.
Similar to the schematic editor, the user can also add/delete points on the trace and change connectivity of points inside connected traces.
Note that the 3D operations maintain the topological connection between pins of parts. These connections are specified in the 2D schematic window and are automatically reflected in the 3D layout window.

%

%
Finally, when the user presses the ``print" button, the system creates the necessary channels and holes on the mesh that correspond to the design, and then exports the geometry for use in 3D printing software.
Channels on the surface are generated using the stroke parametrization technique~\cite{Schmidt13}, which generates texture coordinates with minimum distortion around a stroke.
We simply displace points of the mesh in the normal direction with respect to the distance computed from the parametrization.

\begin{figure}[htbp!]
\includegraphics[width=86mm]{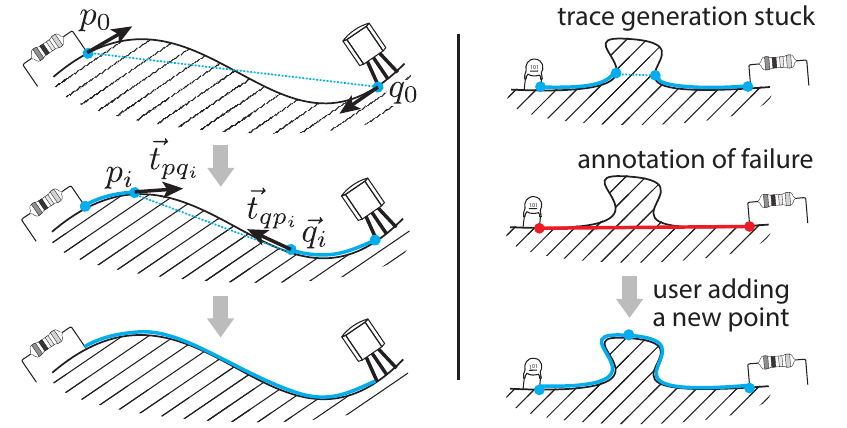}
\caption{Surface trace generation algorithm. (left)~Starting from two endpoints, the 
line segment between the two points is projected on the surface to find tangential
directions for each point. We then incrementally slide each point along the surface in 
that projected direction, until they meet each other. 
(right)~If the algorithm fails we annotate the failure to prompt the user to add an additional point.}
\label{fig:method}
\end{figure}

\begin{figure*}[htbp!]
\includegraphics[width=177.8mm]{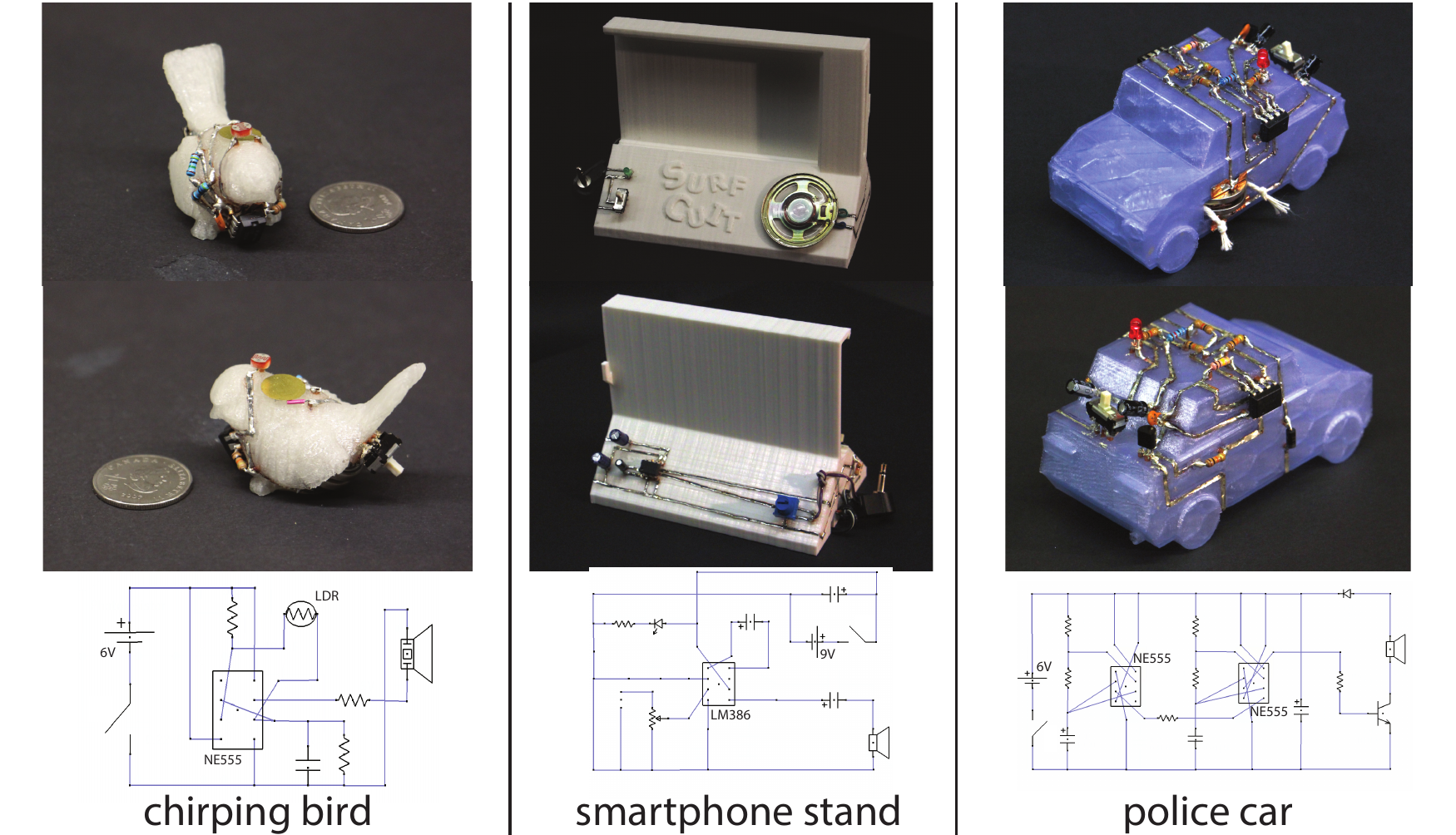}
\caption{Fabricated SurfCuit examples and their schematic diagrams.}
\label{fig:results}
\end{figure*}

\subsection{Algorithmic detail of SurfCuit trace computation}
SurfCuit updates the routing of traces interactively during the user's editing.
A trace is computed as a curve on the surface connecting two end points, each of which is either a pin of a user-specified trace midpoint.
To make manual construction easy, traces should be as short and straight as possible, 
as this will introduce the smallest amount of twisting (torsion) in the copper tape.
The shortest (and thus straightest) path on a surface connecting two points is called a \emph{geodesic}.

There are various existing methods to compute geodesics~(e.g.,~\cite{Surazhsky:2005:FEA:1073204.1073228}), 
however true geodesics are expensive to compute.
Thus, we use a heuristic method to estimate geodesics in real time (see Fig.~\ref{fig:method}-left).
The basic idea behind our method is to start with the two endpoints $p_0$ and $q_0$, and move them towards eachother until they meet.
The direction of movement is defined by a vector in the tangent plane at each point. 
We compute these directinos by first finding a 3D direction, and then projecting into the tangent plane and normalizing. 
For a pair of points $p_i$ and $q_i$, the tangent-plane directions are:
\begin{eqnarray}
\vec{t}_{pq_i} = \frac{(\bi{I}-\vec{n}_{p_i}\otimes\vec{n}_{p_i})(\vec{q}_i-\vec{p}_i) }{|(\bi{I}-\vec{n}_{p_i}\otimes\vec{n}_{p_i})(\vec{q}_i-\vec{p}_i)|}\\
\vec{t}_{qp_i} = \frac{(\bi{I}-\vec{n}_{q_i}\otimes\vec{n}_{q_i})(\vec{p}_i-\vec{q}_i) }{|(\bi{I}-\vec{n}_{q_i}\otimes\vec{n}_{q_i})(\vec{p}_i-\vec{q}_i)|}
\end{eqnarray}
where $\bi{I}$ is an identity matrix, and $\vec{n}_p$ and $\vec{n}_q$ are unit normal vectors at $\vec{p}$ and $\vec{q}$.
We then update the positions of points ($p_i\rightarrow p_{i+1},q_i\rightarrow q_{i+1}$) 
by taking small steps in the directions of $\vec{t}_{pq_i}$ and $\vec{t}_{qp_i}$, respectively.
To keep the resulting point on the triangle mesh, we use discrete parallel transport~\cite{Bergou:2008:DER:1360612.1360662}.
First, a point on a triangle is updated in the direction of the triangle's tangent vector $\vec{t}$ until it hits the boundary (edge) of the triangle. 
Then, at the boundary we transform $\vec{t}$ into the plane of the next triangle using a minimal rotation
around the edge of the triangle, i.e. the rotation that takes the current triangle normal to the neighboring triangle normal.
We stop this update when the $p_i$ and $q_i$ move 1~mm along the surface, and call them $p_{i+1}$ and $q_{i+1}$.

This simple algorithm can fail when the normal of the surface (i.e., $\vec{n}_p$ or $\vec{n}_q$) becomes 
parallel to $\vec{p}_i-\vec{q}_i$~(see Fig.\ref{fig:method}-right).
In such cases, we annotate failure as a line connecting $p_0$ and $q_0$ in order to prompt the user to add another point along the trace. 
Clearly this algorithm does not produce a bounded approximation to a geodesic, however it does produce exact geodesics
for simple surface geometries such as planes or spheres.
And in practice, we have found that it is highly effective at producing low-torsion 3D curves which are ideal for trace fabrication.

%% file: result.tex
\section{SurfCuit Examples}
To demonstrate the effectiveness of our approach, we present seven different examples created using our SurfCuit design tool and fabrication method.
Each example is chosen to showcase various properties of SurfCuit.
Specifically, we demonstrate integration of many different sensors (for light, touch, and sound), controller ICs, and transducers (for light, electromagnetic actuators, sound, and radio wave) into small spaces.
Note that all the examples are fully self-contained and work without external controllers or power sources.
Please refer to the accompanying video for more detail.

We can fabricate SurfCuits on a wide variety of input meshes.
In these examples, the input meshes were taken from the Thingiverse\texttrademark 3D model repository (http://www.thingiverse.com).
We did not need to specifically design new shapes from scratch to accommodate electric circuits.

\subsection{Enriching 3D Prints with Sound and Lights}

\subsubsection{Christmas Tree}
Figure~\ref{fig:teaser} shows a Christmas tree~(\href{https://www.thingiverse.com/thing:608606}{thing:608606}) that blinks 13 LEDs in an asynchronized timing using a 16 pin timer IC, CD4060. 
This example has many components on a relatively small volume. 
21 electric parts, 20 traces, and one 9-volt battery are integrated into the volume (12cm~x~6cm~x~6cm). 
This example also demonstrates the inclusion of an IC with many pins using SurfCuits.
Such complex circuits typically need cables over the traces when constructed on a single-sided 2D board.
However, in the Surfcuit, we can take advantage of the three dimensional structure of the object to avoid such cables.
For example, if we cannot connect two parts on the front side without intersection, the trace can go around the back side.
Because the circuit construction is three-dimensional, there are more degrees of freedom in the trace layout.

\subsubsection{Chirping Birds}
SurfCuit enables the fabrication of complex circuits in a very small volume.
The chirping bird~(Fig.~\ref{fig:results}) integrates a light theremin circuit, which uses a 555 timer IC and photoresistor, into a 3D bird shape (\href{https://www.thingiverse.com/thing:359531}{thing:359531}).
The light theremin circuit modulates the pitch of the sound according to the intensity of light received by an LDR sensor. 
Thus, a user can create chirping sounds by waving a hand on the top of the bird.
This behavior significantly enhances the static bird geometry-- not only does it generates sounds, it makes the bird a playable instrument.
This example also demonstrates circuit integration into a small space. 
The part volume is very small (2.5cm~x~3cm~x~6cm) but because we use the full 3D space we can
fit both the thermemin circuit and 2cm-diameter batteries.
Our interactive layout tool allowed us to avoid obstructing semantically-important features such as the 
bird's face, and place the batteries and switches in the occluded area behind the tail.

\subsubsection{Smartphone Stand}
The speaker-embedded iPhone stand~(Fig.~\ref{fig:results}) augments a smartphone stand shape (\href{https://www.thingiverse.com/thing:642881}{thing:642881}) by integrating a circuit using an LM386 timer IC to amplify the sound signal.
This smartphone stand exemplifies the integration of geometric and electrical functionality. 
While the original geometry provides the function to hold a smartphone, the circuit amplifies the audio signal. 
3D printing makes it easy to fabricate the precise shape needed.  
Achieving the same geometrical functionality is very difficult with flat circuits.

\subsubsection{Police Car}
The police car example~(Fig.~\ref{fig:results}) integrates a circuit that blinks a LED beacon while a magnetic speaker generates siren sounds, which is modulated by two 555 timer ICs.
Aside from the functionality of making sounds and lights of a police car, the surface mounted circuit also gives mechanical appearance to the 3D printed shape. 
The input mesh (\href{https://www.thingiverse.com/thing:806770}{thing:806770}) is clearly a car but it lacks any detailed texture, in large part because the printer is limited to a single material.
The circuitry on the car body gives the shape some definition, creating a more interesting
machine-like appearance that would not be possible with 3D printing alone.

\subsection{Dynamic 3D Prints with High-Current Circuits}
As previously discussed, our traces have only small amounts of parasitic electric resistance, and thus can handle large amount of electric current (up to 1~Ampere, possibly more).
This is enough current drive electro-magnetic actuators, allowing us to create mechanized objects.

\subsubsection{Octopus Fan}
\begin{figure}[htbp!]
\centering
\includegraphics[width=86mm]{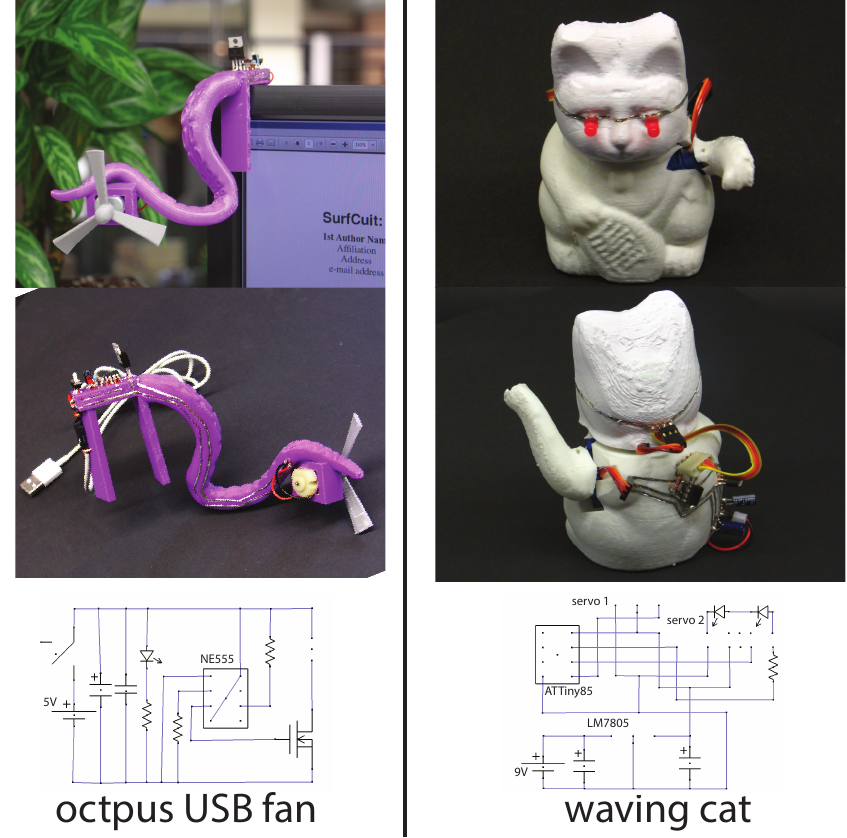}
\caption{Examples of high-current electric circuits. (Left) Octopus USB fan with a touch switch. (Right) Cat robot waving its hand and its head.}
\label{fig:current}
\end{figure}

Our Fan example (Fig.~\ref{fig:current}-left) demonstrates a touch-sensitive USB fan, 
where the user can toggle a DC motor fan on and off by touching specific locations of the print.
The shape of the fan is designed to be clipped to the top of a computer monitor.
Upon making physical contact, a 555 timer IC detects a small current transmitted through body, and toggles the motor control.
We use a MOSFET to amplify the output and drive the 5V DC motor with about 1.0~A current. 
SurfCuit is convenient for fabricating touch-sensitive objects since the traces are naturally exposed to the surface.
We successfully mounted long curved traces on the octopus's tentacle (\href{https://www.thingiverse.com/thing:158069}{thing:158069}).

\subsubsection{Cat Robot}
In the waving cat example~(Fig.~\ref{fig:current}-right), an ATtiny85, which is an Arduino-compatible programmable micro controller, drives two servo motors which wave the arm and shake the head of a cat statue (\href{https://www.thingiverse.com/thing:163032}{thing:163032}).
This robot also draw several hundreds milliamperes of current at peak load. 
The use of a programmable micro-controller makes the interaction/behavior design of this object much more flexible. 
Again, SurfCuit’s highly conductive traces are critical to allowing the micro-controller to drive the various outputs.

\begin{figure}[b!]
\centering
\includegraphics[width=86mm]{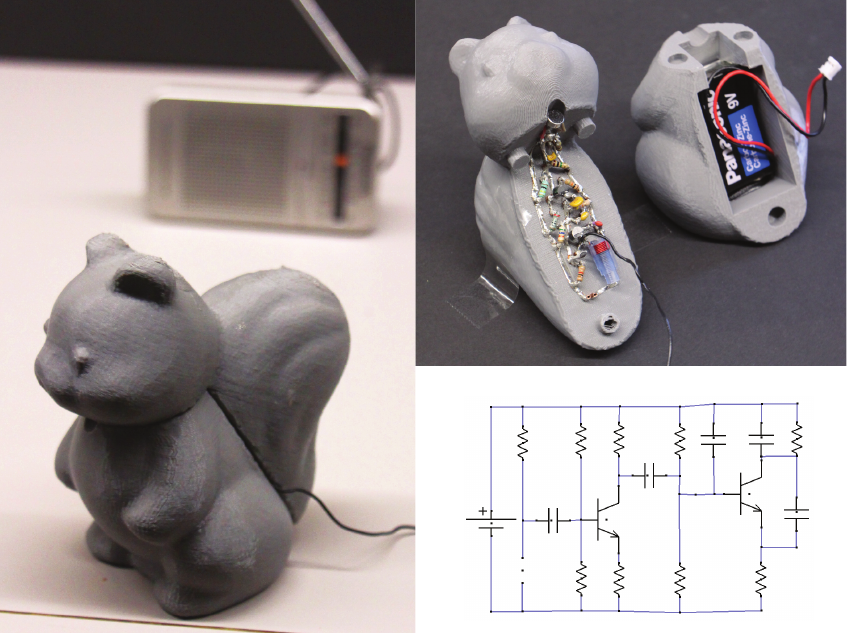}
\caption{A covert “Squirrel Spy” which contains a concealed FM transmitter. }
\label{fig:covert}
\end{figure}

\subsection{Concealed Circuit}
So far, our examples have placed the circuit components and trances on the exterior surface of objects.
These circuits have increased the functionality and interactivity of the 3D prints via sensor-controlled transducers.
However, adding on-surface circuits does involve modifying the original surface, which may be undesirable if the surface has specific functional or aesthetic purposes.
An example of such a requirement arises in cases where we may wish to obscure the functionality of the circuit.
Fig.~\ref{fig:covert} shows a FM transmitter that is concealed inside a shape of a squirrel~(\href{http://www.thingiverse.com/thing:11705}{thing:11705}).
This “squirrel spy” could be used in covert recording or “nanny-mic” type applications.
To create this object, we split the squirrel geometry along a curved partition surface, and then implemented the circuit on the interior curved cross-sections.
Using a curved partition, rather than a planar cut, allows us to design a parting line following concave regions on the surface, which are more easily concealed.
We could of course created a larger cavity and installed a flat PCB board, but with SurfCuit we do not have to worry about the complexities of orienting and mounting the board.
In this example, we also show that SurfCuit can handle very high-frequency circuits (about 100 MHz).
Operating at such frequencies is very sensitive to the parasitic resistance that would be present with less robust fabrication processes.

\subsection{Circuit as a Design Element}
While in some cases we might wish to hide the circuit, in others we can actively use the circuit as part of the design aesthetic.
Many people find beauty in circuits, as seen in the circuit jewelry (e.g., Circuit Breaker Labs~\cite{breaker}) and wearable fashion shows. 
In fact, many ground-breaking works of industrial design have integrated the internal engineering mechanisms 
into aesthetics of the design.
Notable examples include Swiss watches showing the gear work or “movement”, the iMac G3's translucent body, and the intentionally-exposed functional and structural elements of the Pompidou museum in Paris.
Although we cannot claim our own results as works of art, SurfCuit enables this aesthetic by making it easy for creative users to integrate circuitry elements into the external design of complex 3D shapes.
For example, our featureless car above was turned into a police car with more interesting “steampunk” styling.

\begin{figure}[htbp!]
\centering
\includegraphics[width=86mm]{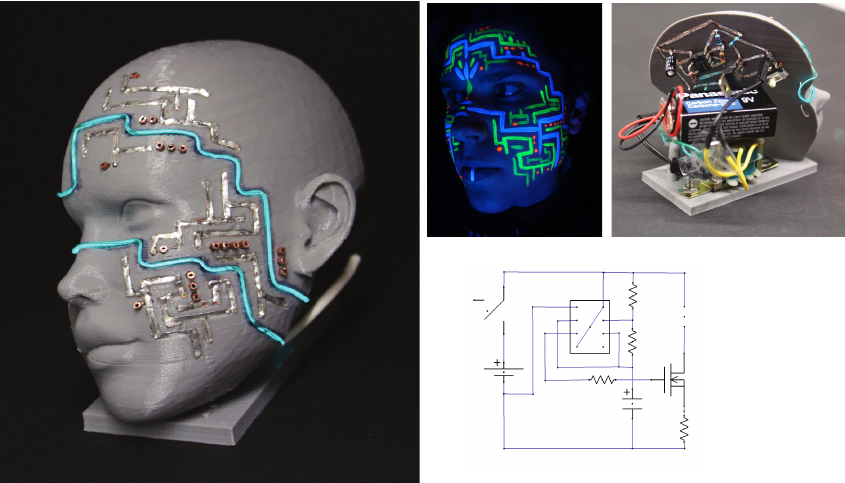}
\caption{ Facial tattoo model uses circuit's trace patterns as a design element (Left). This model is inspired by an artistic circuit tattoo by \href{http://faeriegem.deviantart.com/art/UV-Circuitboard-Face-Paint-255440051}{Faeriegem} (Right). }
\vspace{+9mm}
\label{fig:beauty}
\end{figure}

To further illustrate this concept, we created a circuit that illuminates EL (electro-luminescent) wires whose placement is designed based on an existing circuit-like facial tattoo (see Fig.~\ref{fig:beauty}).
The core of the circuit is an inverter that converts 9V DC current to 120V AC current using 555 timer IC and a micro transformer.
Although most of the traces do not contribute the function of the circuit, the texture of the metal traces completely changes the aesthetic of the otherwise smooth and monochrome 3D-printed head (\href{http://www.thingiverse.com/thing:33503}{thing:33503}).
This example also demonstrates the use of quite high-voltage circuits with SurfCuit,
where insulation between traces is critical.

